\documentstyle[12pt]{article}
\newlength{\extraspace}
\newlength{\extraspaces}
\newcommand{\be}{\begin{equation}
\addtolength{\abovedisplayskip}{\extraspaces}
\addtolength{\belowdisplayskip}{\extraspaces}
\addtolength{\abovedisplayshortskip}{\extraspace}
\addtolength{\belowdisplayshortskip}{\extraspace}}
\newcommand{\ee}{\end{equation}}

\newcommand{\ba}{\begin{eqnarray}
\addtolength{\abovedisplayskip}{\extraspaces}
\addtolength{\belowdisplayskip}{\extraspaces}
\addtolength{\abovedisplayshortskip}{\extraspace}
\addtolength{\belowdisplayshortskip}{\extraspace}}
\newcommand{\ea}{\end{eqnarray}}

\newcommand{\nonu}{\nonumber \\[.5mm]}
\newcommand{\A}{&\!\!\!}

\newcommand{\newsection}[1]{
\vspace{7mm} \pagebreak[3] \addtocounter{section}{1}
\setcounter{subsection}{0} \setcounter{footnote}{0}
\begin{center}
{\large {\bf \thesection. #1}}
\end{center}
\nopagebreak
\medskip
\nopagebreak \hspace{3mm}}

\begin{document}
\pagenumbering{roman}

\newpage

\pagenumbering{arabic}

\begin{center}
{{\bf Kerr-Newman Solution and Energy in Teleparallel Equivalent
of Einstein Theory}}
\end{center}
\centerline{ Gamal G.L. Nashed}

\bigskip

\centerline{{\it Mathematics Department, Faculty of Science, Ain
Shams University, Cairo, Egypt }}

\bigskip
 \centerline{ e-mail:nashed@asunet.shams.edu.eg}

\hspace{2cm}
\\
\\
\\
\\
\\
\\
\\
\\

An exact  charged axially symmetric  solution  of the coupled
gravitational and electromagnetic fields  in the teleparallel
equivalent of Einstein theory is derived. It is characterized by
three parameters ``$\,$the gravitational  mass $M$, the charge
parameter $Q$ and the rotation parameter $a$" and its associated
metric gives  Kerr-Newman spacetime. The parallel vector field and
the electromagnetic vector potential are axially symmetric. We
then, calculate the total energy  using the gravitational
energy-momentum. The energy  is found to be shared by its interior
as well as exterior. Switching off the charge parameter we find
that no energy is shared by the exterior of the Kerr-Newman black
hole.

\newpage
\begin{center}
\newsection{\bf Introduction}
\end{center}

The search for a consistence expression for the gravitating energy
and angular momentum of a self-gravitating distribution of matter
is undoubtedly a long-standing problem in general relativity. The
gravitational field does not possess the proper definition of an
energy momentum tensor and one usually defines some
energy-momentum and angular momentum as Bergmann \cite{BT} or
Landau-Lifschitz \cite{LL} which are pseudo-tensors. The
Einstein's general relativity can also be reformulated in the
context of  teleparallel geometry. In this geometrical setting the
dynamical field quantities correspond to orthonormal tetrad field
${e^i}_\mu$  (i, $\mu$ are SO(3,1) and spacetime indices,
respectively). The teleparallel geometry is a suitable framework
to address the notions of energy, momentum
 and angular momentum of any spacetime that admits a $3+1$
 foliation \cite{ MR}. Therefore, we consider the tetrad theory of
 gravitation.

The tetrad theory  of gravitation  based on the geometry of
absolute parallelism \cite{PP}$\sim$\cite{AGP1} can be considered
as the closest alternative to general relativity, and it has a
number of attractive features both from the geometrical and
physical viewpoints. Absolute parallelism is naturally formulated
by gauging spacetime translations and underlain by the
Weitzenb$\ddot{o}$ck geometry, which is characterized by the
metricity  condition and by the vanishing of the curvature tensor
(constructed from the connection of the Weitzenb$\ddot{o}$ck
geometry). Translations are closely related to the group of
general coordinate transformations which underlies general
relativity. Therefore, the energy-momentum tensor represents the
matter source in the field equation for the gravitational field
just like in general relativity.

The tetrad formulation of gravitation was considered by M\o ller
in connection with attempts to define the energy of gravitational
field \cite{Mo8,Mo2}. For a satisfactory description of the total
energy of an isolated system it is necessary that the
energy-density of the gravitational field is given in terms of
first- and/or second-order derivatives of the gravitational field
variables. It is well-known that there exists no covariant,
nontrivial expression constructed out of the metric tensor.
However, covariant expressions that contain a quadratic form of
first-order derivatives of the tetrad field are feasible. Thus it
is legitimate to conjecture that the difficulties regarding the
problem of defining the gravitational energy-momentum are related
to the geometrical description of the gravitational field rather
than are an intrinsic drawback of the theory \cite{Mj,MDTC}.

It is well known that teleparallel equivalent of general
relativity (TEGR) \cite{HS1} $\sim$ \cite{OP} provides an
alternative description of Einstein's general relativity. In this
theory the gravitational field is described by the tetrad field
${e^a}_\mu$. In fact the first attempt to construct a theory of
the gravitational field in terms of  a set of four linearly
independent vector fields in the Weitzenb$\ddot{o}$ck geometry is
due to Einstein \cite{Wr,Ea,Ea1}.

A well posed and mathematically consistence expression for the
gravitational energy has been developed \cite{MDTC}. It arises in
the realm of the Hamiltonian formulation of the TEGR \cite{MR} and
satisfies several crucial requirements for any acceptable
definition of gravitational energy. The gravitational
energy-momentum $P^a$ \cite{MDTC,MFC} obtained in the framework of
the TEGR has been investigated in the context of several distinct
configuration of the gravitational filed. For asymptotically flat
spacetimes $P^0$ yields the ADM energy \cite{ADM}.

Kawai et. al. \cite{KT} assuming the tetrad field to have the form
\[ {e^i}_\mu={\delta^i}_\mu+\displaystyle{a \over
2}l^kl_\mu-\displaystyle{Q^2 \over 2}m^km_\mu,\] and \[ \eta^{\mu
\nu}l_\mu l_\nu=0, \qquad \qquad \eta^{\mu \nu}m_\mu m_\nu=0,
\qquad \qquad  \eta^{\mu \nu}l_\mu m_\nu=0, \] were able to obtain
a charged Kerr metric solution in new general relativity. In
extended new general relativity also Kawai et. al. \cite{KST} have
examined axi-symmetric solutions of the gravitational and
electromagnetic field equations in vacuum from the point of view
of the equivalence principle.

According to the above discussion, it is clear that there is a
problem in using the definitions of the energy-momentum complexes
resulting from general relativity theory of gravitation. It is the
aim of the present work to derive an exact charged axially
symmetric solution in TEGR for the coupled gravitational and
electromagnetic fields.  Using this solution we calculate the
energy using the gravitational energy-momentum. In \S 2 we derive
the field equations of the coupled gravitational and
electromagnetic fields. In \S 3 we first apply the tetrad field
with sixteen unknown functions of $\rho$ and $\phi$ to the derived
field equations. Solving the resulting partial differential
equations, we obtain an exact analytic solution. In \S 4 we
calculate the energy.  The final section is devoted to discussion
and conclusion.
\newsection{The  TEGR for gravitation and electromagnetism}

In a spacetime with absolute parallelism the parallel vector field
${e_a}^\mu$ define the nonsymmetric affine connection \be
{\Gamma^\lambda}_{\mu \nu} \stackrel{\rm def.}{=} {e_a}^\lambda
{e^a}_{\mu, \nu}, \ee where $e_{a \mu, \nu}=\partial_\nu e_{a
\mu}$\footnote{spacetime indices $\mu, \ \ \nu, \cdots$ and
SO(3,1) indices a, b $\cdots$ run from 0 to 3. Time and space
indices are indicated to $\mu=0, i$, and $a=(0), (i)$.}. The
curvature tensor defined by ${\Gamma^\lambda}_{\mu \nu}$ is
identically vanishing, however. The metric tensor $g_{\mu \nu}$
 is given by
 \be g_{\mu \nu}= \eta_{a b} {e^a}_\mu {e^b}_\nu. \ee

  The Lagrangian density for the gravitational field in the TEGR,
  in the presence of matter fields, is given by\footnote{Throughout this paper we use the
relativistic units$\;$ , $c=G=1$ and $\kappa=8\pi$.}
\cite{MDTC,MVR} \be  {\cal L}_G  =  e L_G =- \displaystyle {e
\over 16\pi}  \left( \displaystyle {T^{abc}T_{abc} \over
4}+\displaystyle {T^{abc}T_{bac} \over 2}-T^aT_a
  \right)-L_m= - \displaystyle {e \over 16\pi} {\sum}^{abc}T_{abc}-L_m,\ee
where $e=det({e^a}_\mu)$. The tensor ${\sum}^{abc}$ is defined by
\be {\sum}^{abc} \stackrel {\rm def.}{=} \displaystyle{1 \over
4}\left(T^{abc}+T^{bac}-T^{cab}\right)+\displaystyle{1 \over
2}\left(\eta^{ac}T^b-\eta^{ab}T^c\right).\ee $T^{abc}$ and $T^a$
are the torsion tensor and the basic vector field  defined by \be
{T^a}_{\mu \nu} \stackrel {\rm def.}{=}
{e^a}_\lambda{T^\lambda}_{\mu
\nu}=\partial_\mu{e^a}_\nu-\partial_\nu{e^a}_\mu, \qquad \qquad
T^a \stackrel {\rm def.}{=} {{T^b}_b}^a.\ee  The quadratic
combination $\sum^{abc}T^{abc}$ is proportional to the scalar
curvature $R(e)$, except for a total divergence term \cite{Mj}.
$L_m$ represents the Lagrangian density for matter fields.  The
electromagnetic Lagrangian density ${\it L_{e.m.}}$ is given by
\cite{KT} \be {\it L_{e.m.}} \stackrel {\rm def.}{=}
-\displaystyle{e \over 4} g^{\mu \rho} g^{\nu \sigma} F_{\mu \nu}
F_{\rho \sigma}, \ee where $F_{\mu \nu}$ being given
by\footnote{Heaviside-Lorentz rationalized units will be used
throughout this paper} $F_{\mu \nu} \stackrel {\rm def.}{=}
\partial_\mu A_\nu-\partial_\nu A_\mu$ and $A_\mu$ being the
electromagnetic potential.

The gravitational and electromagnetic field equations for the
system described by ${\it L_G}+{\it L_{e.m.}}$ are the following
 \ba \A \A e_{a \lambda}e_{b \mu}\partial_\nu\left(e{\sum}^{b \lambda \nu}\right)-e\left(
 {{\sum}^{b \nu}}_a T_{b \nu \mu}-\displaystyle{1 \over 4}e_{a \mu}
 T_{bcd}{\sum}^{bcd}\right)= \displaystyle{1 \over 2}{\kappa} eT_{a
 \mu},\nonu
  \A \A  \partial_\nu \left( e F^{\mu \nu} \right)=0, \ea
where \[ \displaystyle{ \delta L_m \over \delta e^{a \mu}} \equiv
e T_{a \mu}.\] It  is possible to prove by explicit calculations
that the left hand side of the symmetric part of the field
equations (7) is exactly given by \cite{MDTC}
 \[\displaystyle{e \over 2} \left[R_{a
\mu}(e)-\displaystyle{1 \over 2}e_{a \mu}R(e) \right]. \]

\newsection{ Exact Analytic Solution}

Let us begin with the tetrad field  which
 can be written in the spherical polar coordinates as

 \be \left({e^i}_{  \mu} \right)
= \left( \matrix{ A_1(\rho\; ,\phi) &   A_2(\rho\; ,\phi) &
A_3(\rho\; ,\phi) &  A_4(\rho\; ,\phi)
 \vspace{3mm} \cr  B_1(\rho\; ,\phi) \sin\theta
\cos\phi  & B_2(\rho\; ,\phi) \sin\theta \cos\phi & B_3(\rho\;
,\phi) \cos\theta \cos\phi
 & B_4(\rho\; ,\phi) \sin\phi  \sin\theta \vspace{3mm} \cr
  C_1(\rho\; ,\phi) \sin\theta \sin\phi  & C_2(\rho\; ,\phi)  \sin\theta \sin\phi &C_3(\rho\; ,\phi)
   \cos\theta
\sin\phi & C_4(\rho\; ,\phi)  \cos\phi \sin\theta \vspace{3mm} \cr
 D_1(\rho\; ,\phi) \cos\theta&   D_2(\rho\; ,\phi) \cos\theta & D_3(\rho\; ,\phi)\sin\theta  &
   D_4(\rho\; ,\phi) \cos\theta  \cr }
\right)\; , \ee where $A_i(\rho\; ,\phi)\; ,  \ B_i(\rho\;
,\phi)\; , \ C_i(\rho\; ,\phi) \ \ and \ \  D_i(\rho\; ,\phi)\; ,
\
 i=1 \cdots 4 $ are unknown functions of $\rho$  and  $\phi.$
Applying (8) to the field equations (7) we obtain a set of
nonlinear partial differential equations. Due to the lengthy of
writing these partial differential equations we will just write
the solution that satisfy  these differential equations.
\vspace{.3cm}\\
\underline{The Exact Solution}\\

If the arbitrary functions take the following values \ba A_1 \A
=\A 1-\displaystyle{2M \rho-Q^2 \over 2 \Omega }\; , \quad
A_2=\displaystyle{2M \rho-Q^2   \over 2 \Upsilon}\; ,
 \quad A_3=0\; , \quad A_4=-\displaystyle{(2M \rho-Q^2) a \sin^2 \theta \over 2\Omega}\; , \nonu
 B_1 \A =\A \displaystyle{(2 M \rho-Q^2) \over 2 \Omega }\; ,
\quad B_2=\displaystyle {1 \over 2 \Upsilon \cos \phi} \left(2
\rho \alpha- (2 M \rho-Q^2) \cos \phi \right)\; ,\nonu
  B_3 \A =\A \displaystyle{\alpha \over \cos \phi} \; ,
 \quad B_4=\displaystyle {-2 \beta+\displaystyle { (2 M \rho-Q^2) a \sin^2 \theta \cos\phi
  \over \Omega} \over 2 \sin \phi}\; , \nonu
C_1 \A =\A \displaystyle{ (2 M \rho-Q^2)  \over  2\Omega }\; ,
\quad C_2=\displaystyle {1 \over 2\Upsilon \sin \phi}\left(2\rho
\beta-(2 M \rho-Q^2) \sin \phi \right)\; ,\nonu
  C_3\A =\A \displaystyle{\beta \over \sin \phi}\; , \quad
 C_4=\displaystyle {2 \alpha+\displaystyle { (2 M \rho-Q^2)a \sin^2 \theta \sin\phi
  \over \Omega} \over 2 \cos \phi}\; , \nonu
  D_1 \A =\A \displaystyle{(2 M \rho-Q^2)  \over  2 \Omega }\; ,
\quad D_2=1+\displaystyle{(2 M \rho-Q^2) \over   2 \Upsilon }\;
,\nonu
 D_3 \A=\A -\rho\; , \quad D_4=\displaystyle {(2 M \rho-Q^2) a \sin^2 \theta
  \over 2 \Omega}\; ,
 \ea
 where $\Omega, \ \  \Upsilon, \ \ \alpha, \ \ and \ \ \beta$ are defined by
  \ba \Omega \A =\A \rho^2+a^2 \cos^2\theta\; , \qquad \qquad  \Upsilon=\rho^2+a^2-2M\rho+Q^2
  \; , \nonu
\alpha \A=\A \rho \cos \phi+a \sin\phi \qquad \qquad \beta=\rho
\sin \phi-a \cos \phi\;,  \ea
  $M$, $Q$ and $a$ are the gravitational mass, the charge
  parameter and the angular momentum of the rotating source \cite{Tn,KT}.

   The parallel vector field (8) using solution (9) is axially symmetric in
the sense that it is form invariant under the transformation \ba
\A \A \bar{\phi}\rightarrow \phi+\delta \phi\; , \qquad
\bar{e^{(0)}}_\mu \rightarrow {e^{(0)}}_\mu\; , \qquad
\bar{e^{(1)}}_\mu \rightarrow {e^{(1)}}_\mu \cos\delta
\phi-{e^{(2)}}_\mu \sin\delta \phi\; ,\nonu
\A \A \bar{e^{(2)}}_\mu \rightarrow {e^{(1)}}_\mu \sin \delta
\phi+{e^{(2)}}_\mu \cos \delta \phi \; , \qquad \bar{e^{(3)}}_\mu
\rightarrow {e^{(3)}}_\mu . \ea

 The form of  the antisymmetric electromagnetic
 tensor field  $F_{\mu \nu}$ and the energy-momentum tensor are given by
  \ba
  {\bf F} = \A \A
  \displaystyle{Q  \over 2\sqrt{\pi} \Omega^2} \Biggl\{ \left((a^2\cos^2 \theta-\rho^2) d\rho
  +2\rho a^2 \cos \theta \sin\theta d \theta \right)\wedge dt \nonu
  \A \A+ \left(2 \rho a \cos \theta \sin\theta(\rho^2+a^2) d\theta
  - a  \sin^2\theta(\rho^2-a^2\cos^2\theta)
    d\rho\right) \wedge d\phi \Biggr\}\; ,
\ea
  \ba
{T_1}^1 \A=\A -{T_2}^2=-\displaystyle{Q^2 \over 8 \pi \Omega^2}\;
,\nonu
  {T_3}^3 \A=\A -{T_0}^0 =\displaystyle{Q^2(\rho^2+a^2+a^2\sin^2 \theta) \over
8 \pi \Omega^3} \; , \nonu
{T_0}^3 \A=\A \displaystyle{-{T_3}^0 \over \sin^2
\theta(\rho^2+a^2)}  =\displaystyle{Q^2 a\over 4 \pi \Omega^3} .
  \ea
Solution (9) satisfies the field equations (7) and the associated
metric has the following form \ba ds^2=\A \A \left(\displaystyle{
\Upsilon-a^2\sin^2 \theta \over \Omega}\right) dt^2
-\displaystyle{\Omega \over \Upsilon} d\rho^2 -\Omega d\theta^2-
\displaystyle{\left \{ (\rho^2+a^2)^2-\Upsilon a^2 \sin^2 \theta
\right \} \over \Omega } \sin^2 \theta d\phi^2\nonu
\A \A -2 \displaystyle{(2 M \rho-Q^2) a \sin^2 \theta \over
\Omega} dt d\phi
 \; ,\ea which is the Kerr-Newman black hole in the Boyer-Lindquist coordinate.

The previously obtained solutions (Schwarzschild, Reissner
Nordstr$\ddot{o}$m and Kerr spacetimes) can be generated as
special solutions of the tetrad  (8) using (9) by an appropriate
choice of the arbitrary functions. Now we are going to write the
tetrad field (8) using (9) in the Cartesian coordinate.

Performing the following coordinate transformation \cite{KT} \ba
\A \A t \rightarrow t-M\ln \Upsilon-2M^2 \left(1-\displaystyle{Q^2
\over 4M^2}\right) \int{\displaystyle{d\rho \over \Upsilon}}\; ,
\nonu
\A \A x \rightarrow (\rho \cos \phi+a \sin \phi) \sin \theta\; ,
\nonu
\A \A y \rightarrow (\rho \sin \phi-a \cos \phi) \sin \theta\; ,
\nonu
\A \A z \rightarrow \rho \cos \theta\; , \ea where \be
r=\sqrt{x^2+y^2+z^2}=\displaystyle{\sqrt{\rho^4+a^2\rho^2-a^2z^2}
\over \rho}\; , \ee to the tetrad (8) we obtain\footnote{We will
denote the symmetric part by ( \ ), for example$\;$ , $A_{(\mu
\nu)}=(1/2)( A_{\mu \nu}+A_{\nu \mu})$ and the antisymmetric part
by the square bracket [\ ], $A_{[\mu \nu]}=(1/2)( A_{\mu
\nu}-A_{\nu \mu})$ .} \ba \A  \A {e^{(0)}}_0 = 1-{(2M
\rho-Q^2)\rho^2 \over 2\rho_1}\; , \nonu
 \A \A {e^{(0)}}_\alpha = \left\{   -(n_\alpha-\displaystyle {a \over \rho}
 \epsilon_{\alpha  j 3 }\;  n^j)
 -\displaystyle {a^2 \over \rho^2}  \displaystyle {z \over \rho} {\delta_\alpha}^3
  \right\}{(2M \rho-Q^2)\rho^4  \over 2\rho_1(\rho^2+a^2)}=-{b^{(l)}}_0 \; ,    \nonu
\A  \A  {e^{(l)}}_\beta  ={\delta^l}_\beta+ \Biggl \{x^l x_\beta
-2\displaystyle {a \over \rho} \epsilon_{k 3 (\beta} x^{l)} x^k +
\displaystyle {a^2 \over \rho^2} \left[ {\epsilon_{k}}^{l 3}
\epsilon_{m \beta 3} x^k x^m + 2 z\{\rho x^{(l}-\displaystyle{a
\over \rho} {\epsilon_{k 3}}^{(l} \; x^k \}{\delta_{\beta)}}^3
\right] \nonu
\A \A +\displaystyle{a^4 \over \rho^4}z^2 \delta_{\beta 3}
\delta^{3 l} \Biggr \} {(2M \rho-Q^2) \rho^4 \over 2\rho_1
(\rho^2+a^2)^2} \; , \ea where \be \rho_1=\rho^4+a^2z^2\; , \ee
and  the $\epsilon_{\alpha \beta \gamma}$ are the three
dimensional totally antisymmetric tensor with $\epsilon_{1 2
3}=1$. Therefore, we are interested in this solution
 to calculate its associated energy.
\newsection{Energy  associated with the axially symmetric solution}

Now we are going to calculate the energy content of the tetrad
field  (8) using  (9). Before this let us give a brief review of
the derivation of the gravitational energy-momentum.

Multiplication of the symmetric part of Eq. (7) by the appropriate
inverse tetrad field yields it to have the form \cite{MDTC,MVR}
\be
\partial_\nu \left(-e {\sum}^{a \lambda \nu } \right)=-\displaystyle{e
e^{a \mu} \over 4} \left(4{\sum}^{b \lambda \nu }T_{b \nu \mu }-
{\delta^\lambda}_\mu {\sum}^{bdc}T_{bcd} \right)-4\pi {e^a}_\mu
T^{\lambda \mu}.\ee By restricting the spacetime index $\lambda$
to assume only spatial values then Eq. (19) takes the form
\cite{MDTC} \be \partial_0 \left(e {\sum}^{a 0 j}
\right)+\partial_k\left(e {\sum}^{a k j} \right)=-\displaystyle{e
e^{a \mu} \over 4}\left(4{\sum}^{b c j}T_{b c \mu }-
{\delta^j}_\mu {\sum}^{bcd}T_{bcd} \right)-4\pi e {e^a}_\mu T^{j
\mu}.\ee Note that the last two indices of ${\sum}^{abc}$ and
$T^{abc}$ are anti-symmetric. Taking the divergence of Eq. (20)
with respect to j yields

\be -\partial_0 \partial_j \left(-\displaystyle {1  \over 4 \pi} e
{\sum}^{a 0 j} \right)=-\displaystyle{1 \over 16 \pi}
\partial_j\left[e e^{a \mu}\left(4{\sum}^{b c j}T_{b c \mu }-
{\delta^j}_\mu {\sum}^{bcd}T_{bcd} \right)-16 \pi( e {e^a}_\mu
T^{j \mu})\right].\ee

In the Hamiltonian formulation of the TEGR
\cite{MR},\cite{BN}$\sim$\cite{BN2} the momentum canonically
conjugated to the tetrad components $e_{aj}$ is given by \[ \Pi^{a
j}=-\displaystyle {1  \over 4 \pi} e {\sum}^{a 0 j},\] and that
the gravitational energy-momentum $P^a$ contained within a volume
$V$ of the three-dimensional spacelike hypersurface is defined by
\cite{MDTC} \be P^a=-\int_{V} d^3 x
\partial_j \Pi^{a j}.\ee If no condition is imposed on the tetrad
field, $P^a$ {\it transforms as a vector under the global
$SO(3,1)$ group}. It describes the gravitational energy-momentum
with respect to observers adapted to ${e^a}_\mu$.  Let us assume
that the spacetime is asymptotically flat. The total
 gravitational energy-momentum is given by \be P^a=-\oint_{S \rightarrow \infty}
 dS_k \Pi^{a k}.\ee The field quantities are evaluated on a
 surface $S$ in the limit $r \rightarrow \infty.$

 Now we are going to apply Eq. (23) to the tetrad field (8) with Eq. (9) to calculate
 the energy content. We perform the calculations in the Boyer-Lindquist
 coordinate. The only required component of ${\sum}^{\mu \nu \lambda}$

 \ba
{\sum}^{0 0 1} \A =\A {-1 \over
4(\rho^2+a^2\cos^2\theta)^3(\rho^2+a^2-a\rho+Q^2)}
\Biggl[4M\rho^6+6Ma^2\rho^4+2Ma^4\rho^2+2MQ^2\rho^4+4MQ^2a^2\rho^2\nonu
\A \A-2\rho^5Q^2 - 4a^2\rho^3Q^2-2a^4\rho Q^2+2M a^2\cos^2
\theta(\rho^4+\rho^2 a^2 \cos^2 \theta+ a^2 Q^2 \cos^2 \theta+ a^4
\sin^2 \theta )\Biggr.] \ea

Further substituting (24) in (23) we obtain \be
P^{(0)}=E=-\oint_{S \rightarrow \infty}
 dS_k \Pi^{(0) k}=-\displaystyle {1  \over 4 \pi} \oint_{S \rightarrow \infty}
 dS_k  e
{e^{(0)}}_0{\sum}^{0 0 j} \cong \left(M-\displaystyle{Q^2 \over 4
\rho}\left[1+\displaystyle{\left\{3a^2+\rho^2 \over a
\rho\right\}} \tan^{-1} \displaystyle{\left(a \over
\rho\right)}\right] \right).\ee As is clear from (25) that the
energy content is shared by both  the interior and exterior of the
Kerr-Newman black hole. The total energy when $\rho \rightarrow
\infty$ gives the ADM (Arnowitt-Deser-Misner) mass.
\newsection{Main results and discussion}

In this paper we have studied the coupled equations of the
gravitational and electromagnetic fields  in the TEGR, applying
the most general  tetrad (8) with sixteen unknown function of
$\rho$ and $\phi$ to the field equations (7). Exact analytic
solution is obtained (9). This solution (9) is a general solution
from which we can generate the other solutions (Schwarzschild,
Reissner-Nordstr$\ddot{o}$m and  Kerr spacetimes) by an
appropriate choice of the arbitrary functions of the tetrad (8).
The tetrad field and the electromagnetic vector potential are
axially symmetric as is clear from Eq. (11). We then transform
tetrad (8) with (9) into the Cartesian coordinate using the
transformation given by (13).

{\it The geometrical framework determined by the tetrad field and
torsion has proven to be suitable to investigate the problem of
defining the gravitational energy-momentum \cite{MDTC}. A
consistent expression developed in the realm of the TEGR shares
many features with the expected definition. In the context of the
tetrad theories of gravity, asymptotically flat spacetime may be
characterized by the asymptotic boundary condition \[ e_{a \mu}
\cong \eta_{a \mu}+\displaystyle{1 \over 2} h_{a \mu}(1/r),\] and
by the condition $\partial_\mu {e^a}_\nu=O(1/r^2)$ in the
asymptotic limit $r\rightarrow \infty$.  An important property of
the tetrad field that satisfy the above condition is that in the
flat spacetime limit we have ${e^a}_\mu(t,x,y,z)={\delta^a}_\mu$
and therefore ${T^a}_{\mu \nu}=0$. Hence for the flat spacetime we
normally consider a set of tetrad field such that ${T^a}_{\mu
\nu}=0$ in any coordinate system. \cite{MDTC,MR,MVR}}

 Using the definition of the torsion tensor
given by Eq. (5) and apply it to the tetrad field (8) using (9) we
can show that the torsion of flat spacetime is vanishing
identically. Therefore, we use the gravitational energy-momentum
given by (23). The use of (23) is not restricted to Cartesian
coordinate. Therefore, we apply (23) to the tetrad field (8) with
(9) and obtain the energy content (25). As is clear from (25) that
the energy content is shared by both the interior and exterior of
the Kerr-Newman black hole. The asymptotic value of (25)  up to
$O(a^4)$ is given by \be E \cong M-\displaystyle{Q^2 \over \rho}
\left(\displaystyle{1 \over 2}+\displaystyle{a^2 \over
3\rho^2}+\displaystyle{a^4 \over 60\rho^4} \right). \ee This
result is consistent with that obtained before \cite{Vs,Vs1} up to
$O(a^2)$. Switching off the rotation parameter, $(i.e., a=0 )$ the
energy associated will be the same as that of
Reissner-Nordstr$\ddot{o}$m metric \cite{Vc,Tp}. Setting  the
charge parameter $Q$ to be equal zero, i.e., in the case of a Kerr
black hole, it is clear from  expression (26) that there is no
energy contained by the exterior of the Kerr black hole and hence
the entire energy is confined to its interior only.

\end{document}